# Surface relief grating near-eye display waveguide design

**WANG HAODONG, MA DONGLIN**[*]

*School of Optical and Electronic Information, Huazhong University of Science and Technology,Wuhan 430000，Hubei，China*
\* *madonglin@hust.edu.cn*

**Abstract:** A near-eye display device (NED) is a visual optical system that places a miniature display in front of the human eye to provide an immersive viewing experience. NEDs have been playing an irreplaceable role in both early military flight applications and today's civil and entertainment applications. In this paper, we propose an easy-to-machine design of a near-eye display based on surface relief grating waveguides, taking into account the experience of previous designs of near-eye displays, the superior performance of the design, and the accuracy level of existing grating processing. The design is designed to meet the requirements of large field of view and large outgoing pupil extension as much as possible. The design is insensitive to the incident angle and achieves a full-field field-of-view angle of 40°, an angular uniformity error of 20% for diffraction efficiency, and an average diffraction efficiency of 80% for the full field of view. Based on the design, the overall simulation of the optical path of the NED device is completed, and the illumination uniformity of the outgoing pupil expansion of the device is analyzed through simulation.

1. Introduction

Augmented Reality (AR) technology is a technology that superimposes virtual images on real-world things for display interaction. Augmented reality augments or expands the real scene by using image information generated by computer technology, allowing users of augmented reality devices to observe both the real scene around them and the computer-generated augmented information [1-4]. Unlike Virtual Reality (VR) technology, users of augmented reality devices obtain virtual augmented information without losing real scene information. 2012 Google Glass augmented reality glasses launched by Google Inc. marked the official entry of augmented reality near-eye display system into the consumer market [5]. Among the various technical solutions for transmitting optical paths in near-eye display devices, the most common ones are Bird Bath, free-form surface [6-7], and optical waveguide [8-12]. Among them, Bird Bath scheme is the more mature scheme and the most adopted scheme in the commercial field. Bird Bath scheme has excellent display effect, but its size is too large compared with ordinary glasses, so it is not generally considered as the future direction of AR glasses. The advantages of the free-form solution are similar to those of the Bird Bath solution, with excellent display effects and higher optical efficiency than the Bird Bath solution, but the free-form process is complex and there are difficulties in mass production, and the size is also larger than that of ordinary glasses. The optical waveguide solution is currently regarded as the most likely solution for the future of AR glasses, because it has a comparable volume with ordinary glasses, which maximally meets the human fantasy of future AR glasses. Optical waveguide technology can be classified into geometric optical waveguide technology and diffractive optical waveguide technology according to waveguide classification, and Lumus launched DK-40 augmented reality glasses based on geometric optical waveguide technology in 2013 [13]. Geometric light waveguide technology is based on the principle of geometric optics, inserting reflector arrays in the waveguide as a technical solution, the advantage of which is good imaging quality, but the disadvantage is the complex preparation process of reflector arrays, difficult to replicate, low yield of finished products and the existence of ghost images.In 2015, Microsoft launched Hololens smart AR glasses based on diffractive light waveguide [14].

Diffractive waveguide technology is based on the diffraction principle of grating, and there are mainly two technical solutions: Surface Relief Grating (SRG) and Volume Hologram Grating (VHG), of which the surface relief grating solution uses photolithography to make the master and uses nanoimprinting technology to replicate, which is simple to replicate and The product yield is high, which has great advantages in mass production, but the disadvantage is that due to the dispersion effect of the grating, the imaging will appear rainbow effect, and the color uniformity is poor. Body holographic grating is prepared by holographic interference exposure technology, body holographic grating can solve the dispersion problem of surface relief grating, but the disadvantage is that the preparation process is complex, the stability of mass production is poor.

In summary, geometric optical technology solutions for Near-eye Display (NED) devices are difficult to solve the problem of small FOV, large volume and poor illumination and imaging. Optical waveguide technology even in the current obvious disadvantages, but due to its emergence and development time is relatively short, it has been in the mainstream of the current consumer market position, it can be expected that optical waveguide technology for NED devices still has a lot of room for improvement and expansion and is expected to become the optimal solution for near-eye display devices. Combined with the existing grating processing accuracy level, as far as possible to meet the large field of view large pupil expansion needs, this paper proposes a surface relief grating waveguide based on the design of near-eye display devices.

## 2. Principle

### 2.1 Near-Eye Display (NED) Optical System

NED optical system as a visual system has a special image generation method and transmission path, and there are some differences in the principle and design requirements with the traditional visual system. For example, in waveguide-based NED devices, the coupled-in region of the waveguide is required to have high diffraction efficiency, and it needs to have a large exit pupil area and uniform illumination distribution. These requirements are to achieve better performance and experience.

The main technical solutions for the implementation of waveguide-based near-eye display systems are geometric optical waveguides and grating waveguides, as shown in Figure 1. Among them, the grating waveguides are classified as surface relief grating and body holographic grating, and the main difference is the type of coupled-in and coupled-out grating microstructure. For all three types of NEDs, the common infrastructure includes microdisplay, collimating lens set, and optical waveguide.

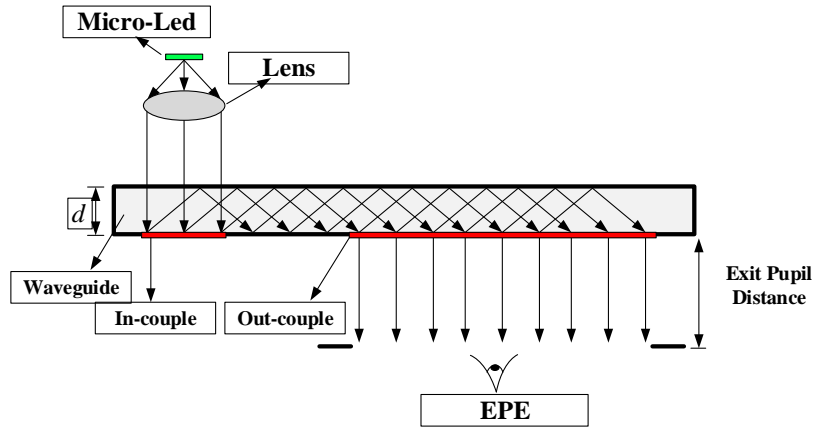

Fig. 1 Waveguide type near-eye display schematic

*2.2 Surface Relief Grating Waveguide*

The surface relief grating has a subwavelength level periodic microsurface structure, and the diffracted beam modulated by it propagates along different diffraction levels and directions, as shown in Figure 2(a), depending on the wavelength of light, the incident angle and the waveguide material. When monochromatic light is considered, the grating period needs to be optimized to obtain a defined diffraction angle. The diffraction efficiency, i.e., the energy ratio of diffracted light at the target level (±1 level), is an important indicator of the energy efficiency of the whole NED system. The diffraction efficiency of the grating with subwavelength periodic structure depends on the slot type and structural parameters of the grating, such as slot depth, tilt angle of the grating, duty cycle, and grating and coating materials, etc. The grating microstructure is shown in Figure 2(b).

The key to the design of surface relief grating NED is to design and optimize the structure of coupled-in and coupled-out gratings. In the coupled-in region, high diffraction efficiency and good angular uniformity are required. In the coupled-out region, it is necessary to achieve pupil expansion and uniform luminance distribution. These elements are crucial for the design of surface relief grating NEDs.

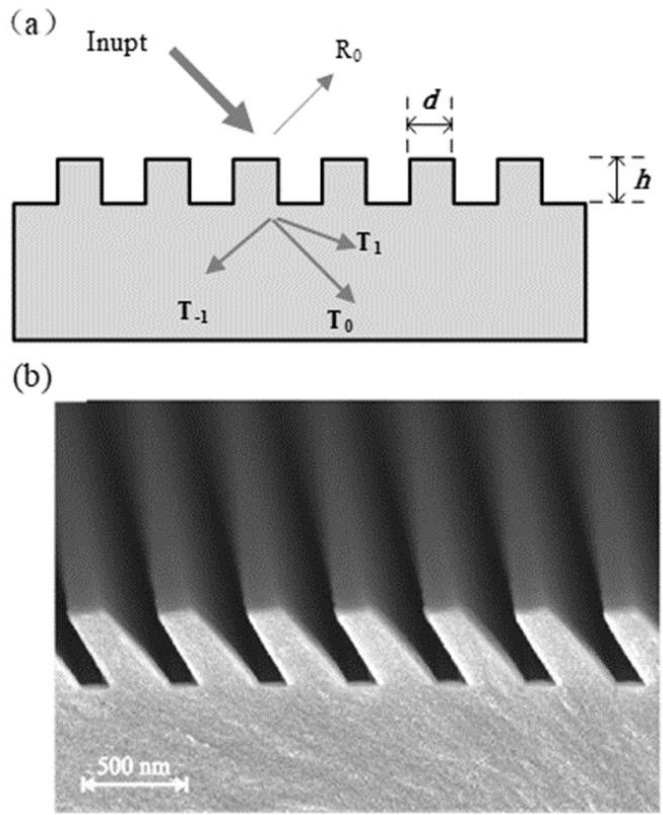

Fig. 2 (a) Surface relief grating schematic; (b) Surface relief grating microstructure diagram

In order to meet the design requirements of large exit pupil area, large field of view, and uniform brightness in the extended exit pupil area, the optical waveguide and grating structure need to be designed accurately. Since the feature size of the surface relief grating is at subwavelength level, the scalar diffraction theory cannot accurately calculate the results, so a rigorous vector diffraction analysis method is required.

By using the vector diffraction analysis method, the diffraction effect of the surface relief grating can be calculated more accurately. This method takes into account the polarization characteristics of light and the details of the waveguide structure, and can provide more accurate results. Through rigorous vector diffraction analysis, the performance of the grating can be evaluated and the structural parameters of the grating can be optimized to achieve brightness uniformity in the large exit pupil region, large field of view and extended exit pupil region as required by the design.

To obtain the maximum diffraction efficiency while reducing stray light from other diffraction stages, the SRG is designed to concentrate the diffracted energy in 1 diffraction stage by the grating diffraction equation

$$\Lambda(\sin i + \sin\theta) = \pm k\lambda \quad (k=1,2,3\cdots) \tag{1}$$

Where $\Lambda$ is the grating period constant, $i$ is the incident angle, $\theta$ is the diffraction angle, k is the diffraction level, and $\lambda$ is the wavelength of the incident light. In order for the diffracted light to satisfy the total reflection condition and propagate in the optical waveguide, the grating period d should be less than the wavelength $\lambda$. In a subwavelength grating, the coupling effect of the components of the electromagnetic field on the boundary surface is not negligible. In this case, the approximate solutions of scalar diffraction theories (e.g. Kirchhoff diffraction theory, Rayleigh-Sommerfeld theory) are no longer applicable. Therefore, a rigorous vector diffraction theory approach is needed to solve the system of Maxwell's equations to obtain accurate diffraction efficiency results for subwavelength gratings.

Rigorous Coupled Wave Analysis (RCWA) is a rigorous vector diffraction theory method proposed by M.G. Moharam and T.K. Gaylord in 1980 [15]. It solves the electromagnetic field vector at each diffraction level by expanding the electromagnetic field vector into coupled wave components at each level, using electromagnetic field boundary conditions at the boundary interface at each level, and by mathematical recursion.

In order to meet the proposed grating design requirements, an analysis of the grating diffraction optical path is required. The expression of the incident light vector K is

$$k_m = \frac{2\pi}{\lambda} n_1 (\sin\theta_0 \cos\varphi_0, \cos\theta_0, \sin\theta_0 \sin\varphi_0) \tag{2}$$

where $\theta$, $\varphi$ are the vector angles in the Cartesian coordinate system, then the vector of the mth diffraction level is

$$k_{i,m} = \frac{2\pi}{\lambda} n_i (\sin\theta_{i,m} \cos\varphi_{i,m}, \cos\theta_{i,m}, \sin\theta_{i,m} \sin\varphi_{i,m}) \tag{3}$$

where i=1 indicates the incident layer and i=0 indicates the reflected layer. For the K-vector incident light, the grating equation is obtained as

$$n_i \sin\theta_{i,m} \sin\varphi_{i,m} = n_1 \sin\theta_0 \sin\varphi_0 = \gamma \tag{4}$$

$$n_i \sin\theta_{i,m} \cos\varphi_{i,m} = n_1 \sin\theta_0 \sin\varphi_0 + m\frac{\lambda}{\Lambda} = \alpha_0 + m\frac{\lambda}{\Lambda} \tag{5}$$

At the same time, the diffracted light needs to satisfy the total reflection angle requirement in the waveguide, which yields

$$\Lambda < \min\left\{\frac{\lambda}{\sqrt{1-\gamma^2}-\alpha_0}, \frac{\lambda}{\sqrt{1-\gamma^2}+\alpha_0}\right\} \tag{6}$$

The grating period $\Lambda$ does not depend on the refractive index of the substrate, and the grating period and the refractive index n of the waveguide material jointly determine the field of view of the NED device. In the grating structure, the polarization mode of the incident light has a significant effect on the diffraction efficiency of the grating. According to the rigorous coupling wave analysis, the TE polarization mode has a higher diffraction efficiency compared to the TM polarization mode under the subwavelength grating condition. Therefore, we choose the TE polarization mode as the incident light.

In the grating coupling, two cases, transmission coupling and reflection coupling, need to be considered. In surface relief grating design, reflection coupling usually achieves higher diffraction efficiency. Therefore, we choose to use reflection-coupled gratings.

To achieve higher reflection diffraction efficiency, a metal substrate is used at the bottom of the grating with a complex refractive index of silver at 525 nm of 0.130+3.159i. The smaller real part and larger imaginary part of silver help to improve the reflection diffraction efficiency of the TE polarization component and absorb more light from the TM polarization component.

To ensure a good uniformity of diffraction efficiency over the full field of view, we coated a titanium dioxide film on the grating surface. The refractive index of titanium dioxide at 525 nm is about 2.985, which is much higher than that of the silver substrate. With the phase-matching condition, the effect of the incident angle on the diffraction efficiency can be significantly reduced, and thus the uniformity of the diffraction efficiency over the full field of view can be obtained.

## 3. Analysis and Discussion

### 3.1 Projection Lens Design

The projection optical lens plays the role of collimating the light beam from the miniature display in this near-eye display device, and in order to achieve the light weight of the whole device and control the cost, fewer and lower cost lenses are used as much as possible. The field of view of the projection lens system is set to 20°×40°.

The optimized projection lens system is shown in Figure 3.

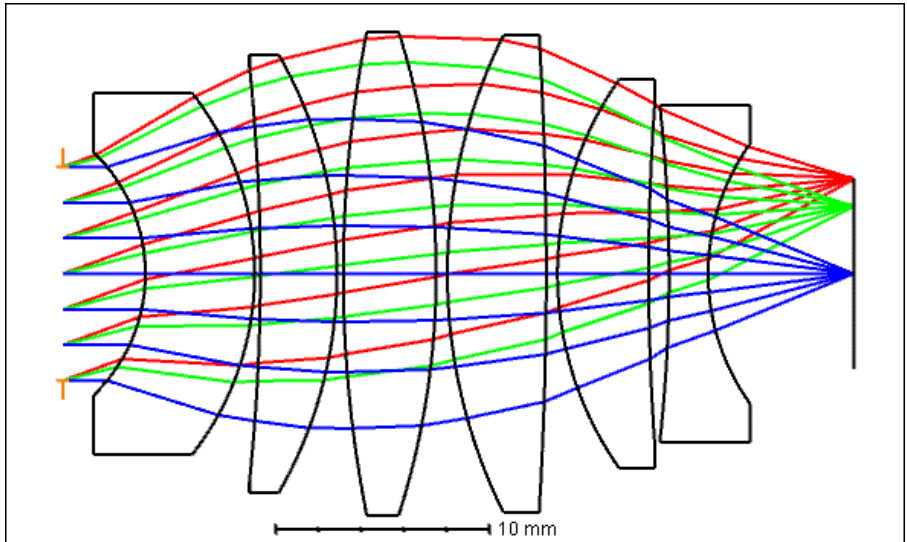
Fig. 3 Optimization of the resulting projection lens system

The total length of the collimation system is 37mm, and the overall length of the system is smaller and the lens size is smaller while obtaining a large field of view with good image quality, which is analyzed as follows:

Modulation Transfer Function (MTF) refers to the modulation system as a function of spatial frequency, which is the most important way to evaluate the imaging performance of the current optical system. Spatial frequency is usually expressed as lp/mm per millimeter line pair, and the modulation system is expressed in the form of luminance contrast of the line pair. As shown in Figure 4, the MTF curves for all fields of view at 20 lp/mm are greater than 0.2, meeting the requirements of the visual optical system.

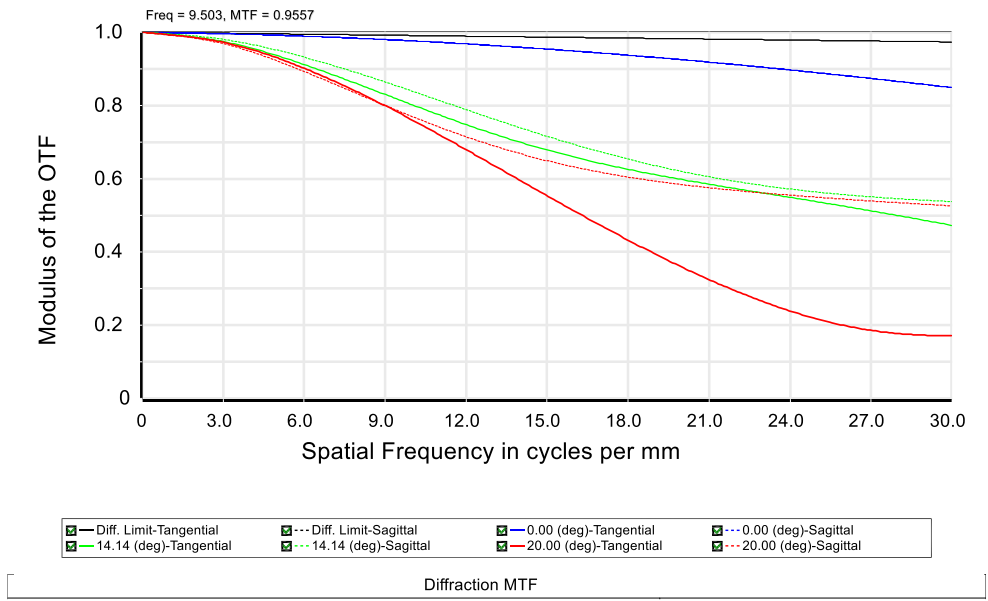
Fig. 4 Modulation Transfer Function curve for each field of view of the projection lens system

*3.2 Diffraction Grating Waveguide Design*

We write the RCWA algorithm in Matlab and use Matlab to calculate and optimize the diffraction grating waveguide. Figure 5 shows the design of a diffractive waveguide with a diagonal field of view of 40° and an aspect ratio of 16:9. The design wavelength is 525 nm, the substrate material is N-LAF2, and the refractive index is 1.74.

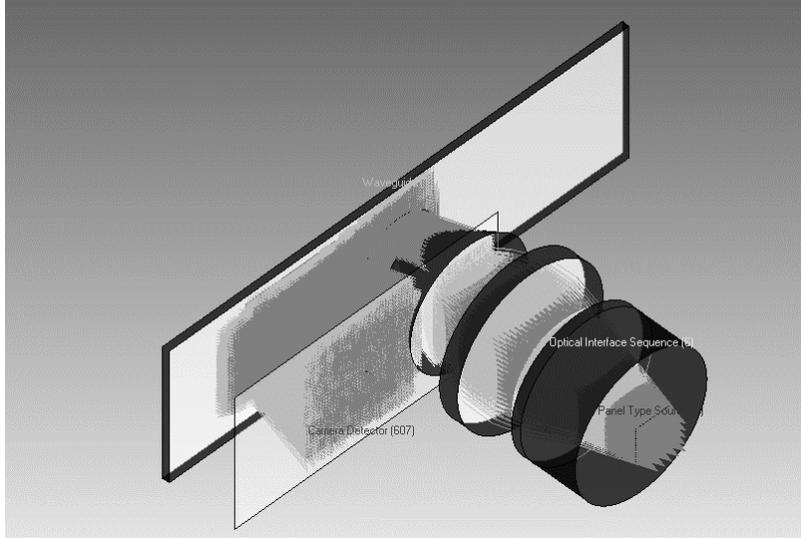

Fig. 5 Surface relief grating waveguide near-eye display system

Matlab software is used to optimize the design of the coupled-in grating with RCWA algorithm. As shown in Fig. 6, the simulation results of the diffraction efficiency of the optimized coupled-in grating in the full field of view and the design wavelength range are shown, where the curve is the optimized diffraction efficiency curve, the diffraction efficiency can reach up to more than 90% at the center wavelength of 525 nm center field of view, the diffraction efficiency of more than 60% at the edge field of view, and the system field of view angle can reach 40°, meanwhile, the average diffraction efficiency reaches 80%, and the uniformity error is 20%

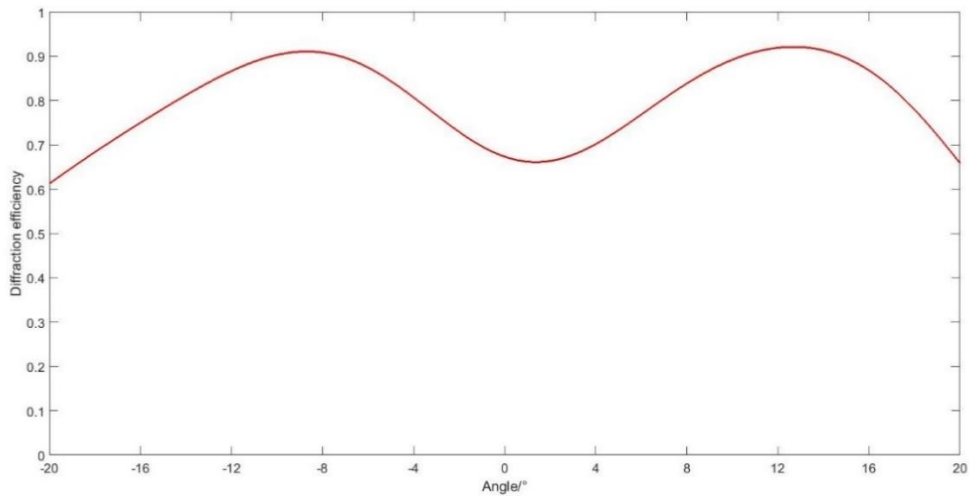

Fig. 6 In-coupling grating diffraction efficiency optimization results

When optimizing the design of the coupled-out grating, the uniformity of energy in the pupil exit region needs to be considered, especially for the illumination uniformity in the extended region of large field of view and large pupil exit. In this design, the optimization of six coupled-out gratings should be closely related.

In practice, the diffraction efficiency of the grating cannot reach 100% and the reflectance cannot reach 0%. In addition, the diffraction efficiency cannot be exactly the same in the full field of view and the full operating band. Therefore, when performing design optimization, it is necessary to reasonably allocate the energy of each grating reflection stage R0 and diffraction stage R-1.

Through reasonable energy allocation, the design optimization of each coupled-out grating in the pupil region can be achieved to achieve better illumination uniformity. Such a design facilitates the subsequent realization of illumination uniformity in the extended region of large field of view and large pupil exit, while meeting the requirements of the NED device input coupler.

When considering the coupled-out grating design, the following strategies can be adopted:

1. The coupled grating 1 is optimized to have a high uniformity of reflected energy R0 and diffracted energy R-1 over the full field of view, which ensures a more uniform beam energy distribution over the full field of view in the coupled region. At the same time, in order to make the corresponding diffraction level energy of the six coupled gratings consistent, the diffraction level energy R-1 of the coupled grating 1 should be reduced and the reflection energy R0 should be increased as much as possible.

2. For the other three coupled-out gratings, the energy distribution of the six coupled-out gratings can be achieved by increasing the diffraction level energy R-1 and decreasing the reflection energy R0, while maintaining the diffraction energy distribution of each field of view as much as possible.

3. It is worth noting that the transmission efficiency of the full field of view should be considered when designing the coupled-out grating, which is related to the ambient light transmission rate of the whole near-eye display system.

However, it should be noted that in practice the light incident on the pupil exit region is not a plane wave of uniform energy. The natural vignetting and edge field illumination attenuation as well as the full-field diffraction efficiency inhomogeneity of the coupled-in grating can affect the design of the coupled-out grating. One of the most important indicators of the imaging performance of NED devices is the uniformity of the extended pupil illumination.

The coupled-in grating and coupled-out grating are coupled into the waveguide, and a comparison of the simulation results of the forward-incidence system before and after the optimization strategy is shown in Figure 7.

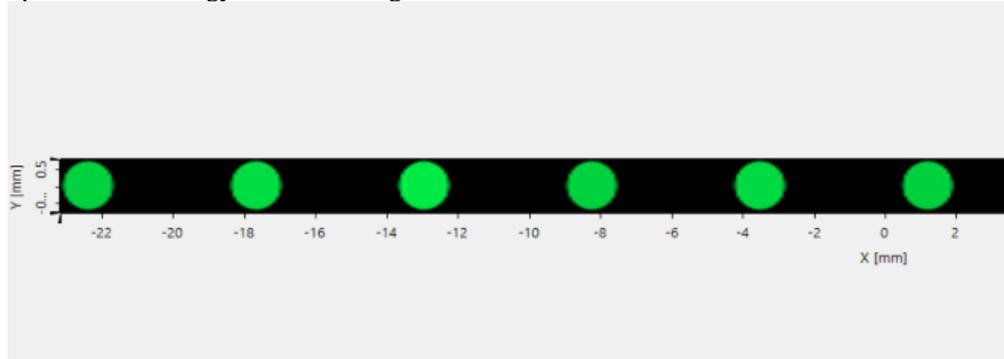

Fig. 7 Display effect of pupil area after optimization of pupil expansion uniformity

After pupil exit uniformity optimization, the simulation results of the full-field incidence system are shown in Figure 8. The illuminance of each region in the eye movement range is 0.044, 0.041 and 0.038 (V/m)2, respectively, with a uniformity error of 12%, which meets the human eye viewing requirements. It is worth pointing out that the main light can be optimized according to the specific structure of the waveguide, which can also achieve a better display effect in the pupil exit area.

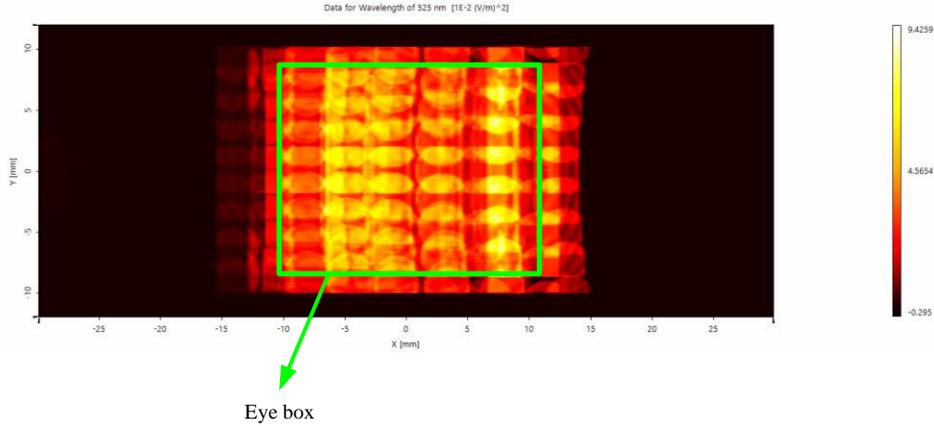

Eye box

Fig. 8 Illumination of the pupil exit area in the full field of view

The final coupled-in and coupled-out grating data obtained after optimization are shown in Table 1. From the data in the table, it can be seen that the ratio of modulation depth to line width is less than 2 for all the gratings in this design, which is highly processable and easy to implement. The diffraction efficiency of the coupled grating is shown in Fig. 9(a), and the transmission rate of the coupled grating is shown in Fig. 9(b), and the transmission rate of the full field of view is above 70%. The waveguide cross section is shown in Fig. 10.

Table. 1 Optimization of the final In-coupling and Out-coupling gratings

| Gratings | Angle/° | Fill factor | Depth/nm | Coating/nm |
|---|---|---|---|---|
| In-couple | 25 | 66% | 415 | 140 |
| Out-couple1 | 0 | 18% | 143 | 166 |
| Out-couple2 | 0 | 18% | 158 | 166 |
| Out-couple3 | 0 | 18% | 179 | 262 |
| Out-couple4 | 0 | 18% | 232 | 221 |
| Out-couple5 | 0 | 18% | 304 | 144 |
| Out-couple6 | 0 | 18% | 384 | 205 |

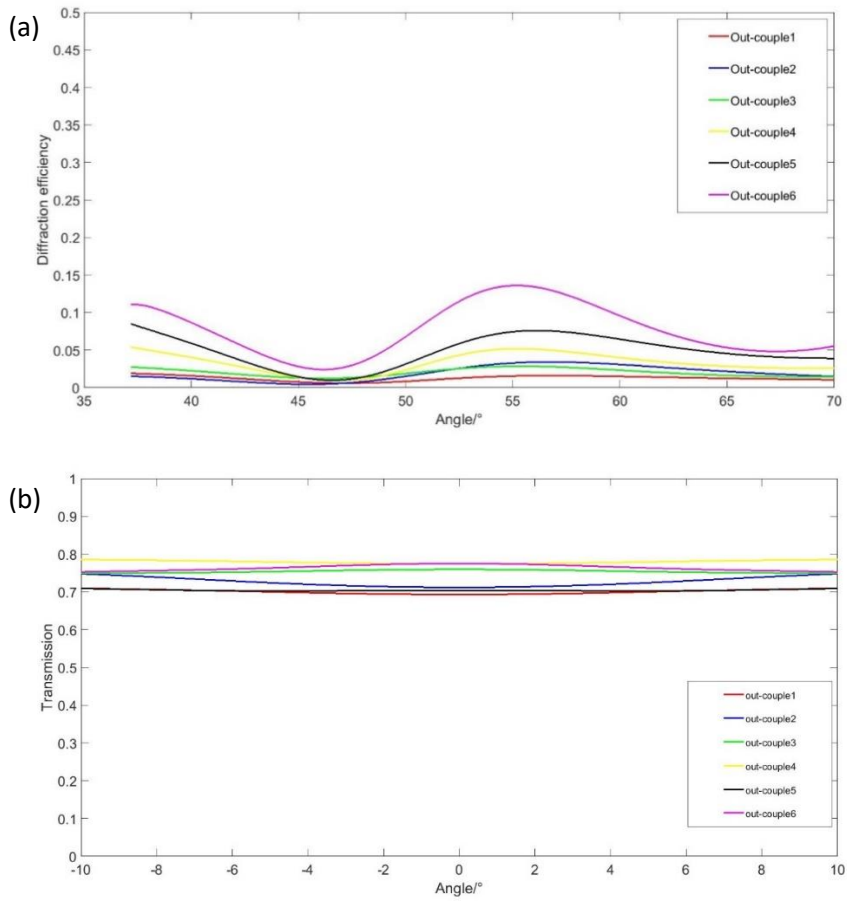

Fig. 9 (a) Out-coupling grating full-field diffraction efficiency; (b) Out-coupling grating full-field transmittance

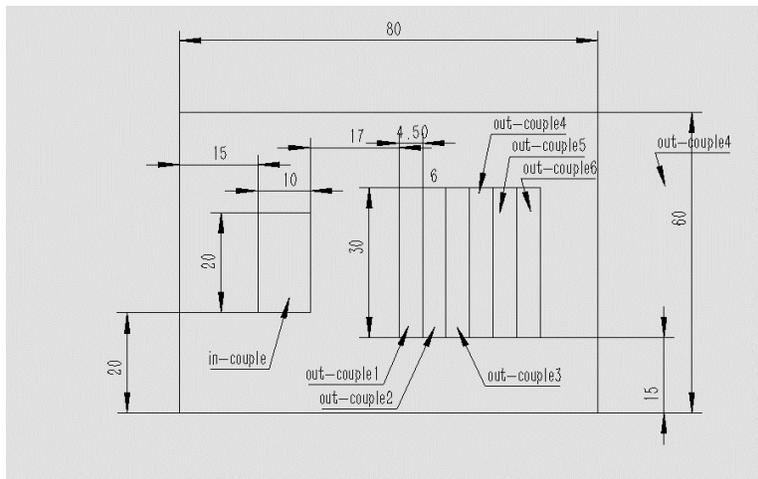

Fig. 10 Waveguide cross-section diagram

*3.3 Design results and testing results*

Now based on our simulation in Zemax as well as VirtualLab, our provided AR module has superior imaging performance. Our built testing facility has validated our design results. The design performance based on software simulation as well as the testing performance can be listed in the following table.

**Table. 2 Simulation performance of our designed AR module**

| Optical quantities | Specifications |
| --- | --- |
| Field of View | >40° |
| MTF @ 18 lp/mm | >0.4 |
| Total weight of AR module | <50 g |
| Distortion | <5% |
| RMS WFE | <0.25λ |
| Magnification testing accuracy | <1.5% |
| MTF testing accuracy for AR module | <±5% |
| Resolution testing | Capable |

## 4. Conclusion

In this paper, we focus on the design and optimization of near-eye augmented reality devices (NEDs), using surface relief lenticular waveguide technology to achieve lightweight, compact, and portable devices with large field of view and extended pupil area, as well as good imaging performance, so that users can enjoy a good augmented reality experience even in motion.

In this study, we propose a near-eye augmented reality device that is compact, portable, easy to process and prepare, and has a large field of view and extended pupil area.

For the design of grating couplers in NED devices, we designed the coupled-in grating and coupled-out grating systems respectively. The initial grating structure design is obtained by rigorous coupling wave analysis, and the coupled-in grating is optimized. We propose a coupled-in grating design that achieves large field-of-view diffraction efficiency uniformity. The design is insensitive to the angle of incidence, with a full field-of-view angle of 40°, and with TE polarization incidence, the angular uniformity of diffraction efficiency reaches 80%, and the average diffraction efficiency of the full field of view reaches 80%.

To meet the requirement of illumination uniformity in the extended pupil area, we propose the design of six coupled-out gratings, and reasonably allocate the reflected energy R¬0 and diffraction level energy R¬-1 for each grating, and optimize the diffraction efficiency of the coupled-out gratings in the pupil area by combining the designed collimation system and the illumination uniformity of the coupled-in gratings.

Finally, we performed the overall simulation of the optical path of the NED device and analyzed the illumination uniformity in the extended pupil region. The simulation results show that after the optimization, the eye movement range of the NED device is 20mm x 16mm, and the illumination uniformity of the extended pupil area is improved from 23% to 88%. At the same time, the imaging quality of the device in the full field of view meets the human eye viewing requirements, and the MTF is greater than 0.4 at 18 lp/mm.